\documentclass[useAMS,usenatbib]{mn2e}
\usepackage{times,amsmath,amssymb,amsfonts,latexsym}
\usepackage{graphicx,fleqn,colordvi}
\newcommand{\dd}{{\rm d}}              \newcommand{\bv}{\mathbf{v}}
\newcommand{\rmth}{{\rm th}}           \newcommand{\dm}{{\rm dm}}

\newcommand{\mnras}{MNRAS}             \newcommand{\aap}{A\&A}

               \newcommand{\physrep}{Phys. Rep.}
\begin{document}
\label{firstpage}
\title[Statistical computation of Boltzmann entropy]{Statistical computation of Boltzmann entropy and estimation of the optimal probability density function from statistical sample}
\author[N. Sui, M. Li and P. He]{Ning Sui$^{1}$, Min Li$^{2}$ and Ping He$^{1,3,4}$\thanks{E-mail: hep@itp.ac.cn}\\
$^1$College of Physics, Jilin University, Changchun 130012, China \\
$^2$Changchun Artificial Satellite Observatory, Chinese Academy of Sciences, Changchun 130117, China \\
$^3$Center for High Energy Physics, Peking University, Beijing 100871, China\\
$^4$State Key Laboratory of Theoretical Physics, Institute of Theoretical Physics, Chinese Academy
of Sciences, Beijing 100190, China}
\date{\today}
\maketitle
\begin{abstract}
In this work, we investigate the statistical computation of the Boltzmann entropy of statistical samples. For this purpose, we use both histogram and kernel function to estimate the probability density function of statistical samples. We find that, due to coarse-graining, the entropy is a monotonic increasing function of the bin width for histogram or bandwidth for kernel estimation, which seems to be difficult to select an optimal bin width/bandwidth for computing the entropy. Fortunately, we notice that there exists a minimum of the first derivative of entropy for both histogram and kernel estimation, and this minimum point of the first derivative asymptotically points to the optimal bin width or bandwidth. We have verified these findings by large amounts of numerical experiments. Hence, we suggest that the minimum of the first derivative of entropy be used as a selector for the optimal bin width or bandwidth of density estimation. Moreover, the optimal bandwidth selected by the minimum of the first derivative of entropy is purely data-based, independent of the unknown underlying probability density distribution, which is obviously superior to the existing estimators. Our results are not restricted to one-dimensional, but can also be extended to multivariate cases. It should be emphasized, however, that we do not provide a robust mathematical proof of these findings, and we leave these issues with those who are interested in them.
\end{abstract}

\section{Introduction}
\label{sec:intro}

Entropy is very important in thermodynamics and statistical mechanics. In fact, it is the key concept, upon which the equilibrium statistical mechanics is formulated \citep{landau96,huang87}, and from which all the other thermodynamical quantities can be derived. Also, the increasing of entropy shows the time-evolutionary direction of a thermodynamical system. The concept of entropy can even be applied to non-thermodynamical systems such as information theory, in which entropy is a measure of the uncertainty in a random variable \citep{shannon48,shunsuke93}, and to non-ordinary thermodynamical systems such as self-gravitating systems, in which the dominating microscopic interaction between particles is long-ranged \citep{lb67,campa09}. The latter is relevant to our studies, in which we formulated a framework of equilibrium statistical mechanics for self-gravitating systems \citep{hep10,hep11,kang11,hep12a,hep12b}. In these works, we demonstrated that the Boltzmann entropy
\begin{equation}
\label{eq:bs}
S_{\rm B}[F] = -\int F(\mathbf{v}) \ln F(\mathbf{v}) \dd^3 \mathbf{v}
\end{equation}
is also valid for self-gravitating systems, in which $F(\bv)$ is the system's probability density function (PDF, hereafter).

Hence, the PDF is necessary for computing the systems' entropy, but often, instead of giving an analytic form of PDF, we have to deal with a thermodynamical system that is in data-form. For instance, \citet{helmi99} performed numerical simulations to study satellite galaxy disruption in a potential resembling that of the Milky Way. In their work, the coarse-grained Boltzmann entropy is used as a measure of the phase-mixing to indicate how mixing of disrupted satellites can be quantified. The analytic PDF is unavailable, and hence they derived the coarse-grained PDF from the simulation data by histogram, that is, by taking a partition in the 6D phase-space and counting how many particles fall in each 6D cell.

\begin{figure}
\centerline{\includegraphics[width=1.0\columnwidth]{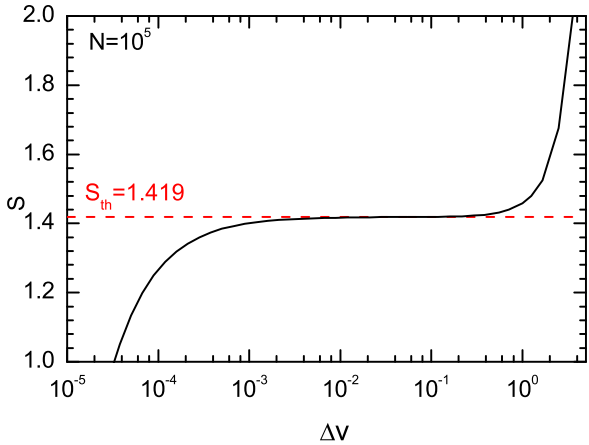}}
\caption{Illustration of the monotonic behaviour of $S(\Delta v)$. The underlying distribution is the univariate standard normal distribution, from which the data set is randomly drawn, with the sample size $N = 10^5$. $\Delta v$ is the bin width of the one-dimensional histogram, and $S$ is evaluated by equation~(\ref{eq:bs}), but with $\hat{F}_{\Delta v}(v)$ instead of $F(v)$. $S(h)$, computed by kernel estimation, exhibits the similar behaviour with respect to the bandwidth $h$. $S_\rmth$, shown with the dashed line, is directly computed by using the analytical univariate standard normal PDF of equation~(\ref{eq:1dgauss}).}
\label{fig:sdv105}
\end{figure}

So, with analytic PDF unavailable, it is indispensable to compute the system's entropy from the data-based samples. This seemingly easy computation, however, is plagued with some unexpected troubles. As we have seen in equation~(\ref{eq:bs}), the evaluation of entropy is actually related to the estimation of probability density, which is one of the most important techniques in exploratory data analysis (EDA). Usually, $\hat{F}_{\Delta\bv} (\mathbf{v})$, the estimation of the underlying unknown PDF $F(\bv)$ of a statistical sample, is derived by data binning, i.e. histogram \citep{gentle09}, just as \citet{helmi99}. However, in the real practice, we find the following interesting phenomenon. See Figure~\ref{fig:sdv105}, the resulting entropy $S$ depends on the bin width $\Delta v$, monotonically increasing with the bin width, and thus it is troublesome to select an appropriate bin width of histogram for computing the entropy.

Prior to our study, there have been many investigations on how to make the optimal probability density estimation of statistical samples \citep{scott92,gentle09}. Histogram is the mostly used method, in which the bin width is its characteristic parameter. A larger bin width produces over-smooth estimation and the PDF looks smoother and featureless, while a smaller bin width produces under-smooth estimation so that the resulting PDF has large fluctuations and spurious bumps. Hence, the selection of an optimal bin width is indeed an important task in EDA, and such an effort can be dated back to \citet{sturges26}, but the famous selector for optimal bin width is given by \citet{scott79}. Usually, the optimal bin width is obtained by minimizing the asymptotic mean integrated squared error (AMISE), which is a tradeoff between the squared bias and variance. Contrary to the conventional treatment, however, \citet{knuth13} proposed a straightforward data-based method of determining the optimal number of bins in a uniform bin-width histogram using Bayesian probability theory.

Another commonly used density estimation method is kernel estimation method, which was first introduced by \citet{parzen62} and \citet{rosenblatt56} and is considered to be much superior to the histogram method \citep{jones96}. With kernel density estimation, we arrive at the similar results as with histogram that the entropy increases as a monotonic function of the kernel bandwidth $h$.

\begin{figure}
\centerline{\includegraphics[width=0.95\columnwidth]{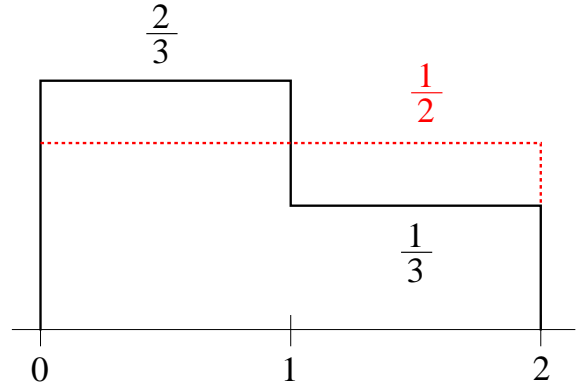}}
\caption{The origin of the monotonicity of $S(\Delta v)$ versus $\Delta v$ is caused by coarse-graining of the PDF. The step solid line and the dotted line indicate the fine-grained and coarse-grained PDF, respectively. The entropy evaluated with coarse-grained PDF is larger than that with fine-grained PDF. See Section~\ref{sec:origin} for detail. This explanation is also applicable to $S(h)$ versus $h$.}
\label{fig:dfcg}
\end{figure}

Besides the two main methods, there are also other methods of density estimation, such as average shifted histogram, orthogonal series estimators \citep{gentle09,scott92,scott04}. We just focus on histogram and kernel methods in this work.

Investigations on histogram and kernel estimation have indeed provided us with useful criteria for optimal bin width of histogram or bandwidth of kernel function \citep{silverman86,jones96,wand97}, yet all these criteria need to take into account the functional form of the true PDF, which is usually unavailable. So it is a difficulty on how to select an optimal bin width or bandwidth for computing the entropy. Fortunately, by large amounts of numerical experiments, we find that the first derivatives of $S(\Delta v)$ with respect to $\Delta v$ for histogram estimation, or of $S(h)$ with respect to $h$ for kernel estimation, correspond to the optimal bin width and bandwidth in all the cases we considered in this work.

In this paper, we describe our numerical experiments and demonstrate this finding in detail. Although we do not provide a robust mathematical proof of this finding, we suggest that the first derivative of entropy be regarded as an alternative selector other than the usual AMISE to pick out an optimal bin width of histogram and bandwidth of kernel estimation. The paper is organized as follows. In Section~\ref{sec:method}, we introduce our methods and describe the working procedure. In Section~\ref{sec:result}, we present the results. Finally, we give our conclusions and discussions in Section~\ref{sec:summy}.

\begin{figure*}
\centerline{\includegraphics[width=2.05\columnwidth]{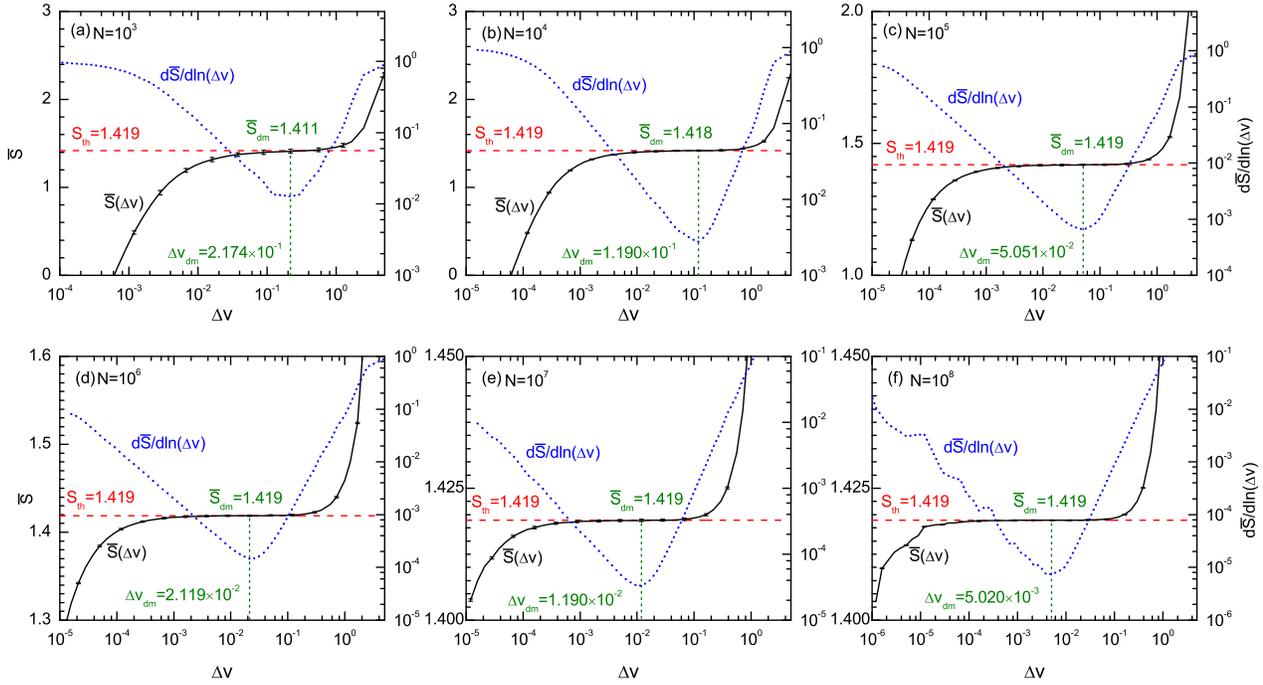}}
\caption{$\bar{S}(\Delta v)$, the averaged entropy evaluated by histogram density estimation and its first derivative w.r.t to $\Delta v$, $\dd \bar{S}/\dd \ln(\Delta v)$, for the one-dimensional standard normal distribution of equation~(\ref{eq:1dgauss}). Error bars of $\bar{S}(\Delta v)$ curves indicate the standard deviation. In every case, the mean and the standard deviation are evaluated with $50$ instances. $\Delta v$ is the bin width of the histogram. In every panel, the vertical axis on the left represents the entropy $S$, and the right represents the first derivative, $\dd \bar{S}/\dd \ln(\Delta v)$. The entropies and their derivatives as functions of $\Delta v$ are indicated as solid and dotted lines, respectively. For the visual clarity, the derivatives are exhibited in logarithmic scale. The subscript `dm' in the figure indicates `derivative's minimum', with $\bar{S}_\dm \equiv S(\Delta v_\dm)$. $S_\rmth$, shown with dashed lines, is the entropy evaluated with analytical PDF of equation~(\ref{eq:1dgauss}). }
\label{fig:1dgaussDv}
\end{figure*}

\section{Methods and procedure}
\label{sec:method}
\subsection{Origin of the monotonicity of $S(\Delta v)$ and $S(h)$}
\label{sec:origin}

We demonstrate the origin of the monotonicity of $S(\Delta v)$ verse $\Delta v$. See Figure~\ref{fig:dfcg}, we design a simple one-dimensional normalized experimental PDF, as:
\begin{eqnarray}
\label{eq:tpdf1}
f_1(v)=\left\{\begin{matrix}
 & \frac{2}{3}, & 0<v<1, \\
 & \frac{1}{3}, & 1<v<2, \\
 & 0,           & {\rm otherwise},
\end{matrix}\right.
\end{eqnarray}
with which the entropy evaluated by equation~(\ref{eq:bs}) is $S_1=\ln 3 - \frac{2}{3}\ln2$. Next, the averaged, or coarse-grained, PDF within the interval $0<v<2$ can be easily derived and normalized as:
\begin{eqnarray}
\label{eq:tpdf2}
f_2(v)=\left\{\begin{matrix}
 & \frac{1}{2}, & 0 < v < 2, \\
 & 0,           & {\rm otherwise}.
\end{matrix}\right.
\end{eqnarray}
So, the entropy evaluated with this coarse-grained PDF $f_2$ is, $S_2=\ln2$, and we can immediately see that $S_2 > S_1$. Hence, the monotonicity of $S(\Delta v)$ verse $\Delta v$ is caused by the coarse-graining of PDF. The same explanation is also applicable to the monotonicity of $S(h)$ versus $h$.

Beyond the monotonicity of $S(\Delta v)$, however, there should be another important feature hiding in the $S(\Delta v)$ curve. By scrutinizing $S(\Delta v)$ of Figure~\ref{fig:sdv105}, we speculate that the first derivative of $S(\Delta v)$ with respect to $\Delta v$, might take its local minimum around the cross point of $S_\rmth$ and $S(\Delta v)$. If this is the case, then we can use this property to construct a selector of optimal bin width of histogram, so that we can compute the entropy to the best extent. We will verify this speculation below.

\begin{table*}
\caption{Results derived from histogram estimation. Listed are the mean $\bar{S}_{\rm dm}$ and the variance $\sigma(\bar{S}_{\rm dm})$ of the entropy at the minimum point, $\Delta {\bf v}_{\rm dm}$, of the derivative of entropy for the three experimental PDFs addressed in Section~\ref{sec:expdf}. The mean and the variance are evaluated from $50$ instances with different sets of random numbers. Sample sizes for the experiments range from $10^3$ to $10^8$. To compare with, the theoretical results $S_{\rm th}$ for the three distributions are 1.419, 0.284, and 4.257, respectively.}
\label{tab1}
\begin{tabular}{cccccccccc}
\hline
  & \multicolumn{3}{c}{1D normal}  & \multicolumn{3}{c}{1D power-law} & \multicolumn{3}{c}{3D normal} \\
N &  $\Delta {\bf v}_{\rm dm}$ & $\bar{S}_{\rm dm}$  & $\sigma(\bar{S}_{\rm dm})$  & $\Delta {\bf v}_{\rm dm}$  & $\bar{S}_{\rm dm}$ & $\sigma(\bar{S}_{\rm dm})$  & $\Delta {\bf v}_{\rm dm}$ & $\bar{S}_{\rm dm}$ & $\sigma(\bar{S}_{\rm dm})$  \\
\hline
$10^3$ &  2.174$\times10^{-1}$ & 1.411  & 2.099$\times10^{-2}$ & 6.250$\times10^{-2}$  & 0.2756 & 9.919$\times10^{-3}$ & 1.170$\times10^{-1}$ &  4.226  & 4.003$\times10^{-2}$ \\
$10^4$ &  1.190$\times10^{-1}$ & 1.418  & 6.643$\times10^{-3}$ & 2.419$\times10^{-2}$  & 0.2790 & 3.121$\times10^{-3}$ & 6.511$\times10^{-2}$ &  4.254  & 1.083$\times10^{-2}$ \\
$10^5$ &  5.051$\times10^{-2}$ & 1.419  & 2.115$\times10^{-3}$ & 8.721$\times10^{-3}$  & 0.2797 & 4.941$\times10^{-4}$ & 3.669$\times10^{-2}$ &  4.256  & 3.697$\times10^{-3}$ \\
$10^6$ &  2.119$\times10^{-2}$ & 1.419  & 8.134$\times10^{-4}$ & 3.178$\times10^{-3}$  & 0.2812 & 2.464$\times10^{-5}$ & 1.546$\times10^{-2}$ &  4.256  & 1.295$\times10^{-3}$ \\
$10^7$ &  1.190$\times10^{-2}$ & 1.419  & 2.045$\times10^{-4}$ & 1.786$\times10^{-3}$  & 0.2805 & 1.326$\times10^{-5}$ & 8.695$\times10^{-3}$ &  4.257  & 4.184$\times10^{-4}$ \\
$10^8$ &  5.020$\times10^{-3}$ & 1.419  & 8.081$\times10^{-5}$ & 1.004$\times10^{-3}$  & 0.2804 & 6.098$\times10^{-7}$ & 1.151$\times10^{-3}$ &  4.257  & 1.049$\times10^{-4}$ \\ \hline
theoretic & -- & 1.419 & -- & -- & 0.2804 & -- & -- & 4.257 & -- \\
\hline
\end{tabular}
\end{table*}

\subsection{Three experimental PDFs}
\label{sec:expdf}

Our strategy is briefly described as follows. First, choose some analytical PDFs, and with Monte Carlo technique, we draw $N$ random data from these analytical PDFs.  With these statistical samples, we can construct the estimators $\hat{F}_{\Delta v}(v)$ or $\hat{F}_h(v)$ of the true PDFs, depending on whether using histogram or kernel estimation. Secondly, we can compute the entropy by equation~(\ref{eq:bs}), with $\hat{F}_{\Delta v}(v)$ or $\hat{F}_h(v)$ replacing the true PDFs. In this way, we derive the curves of $S(\Delta v)$ or $S(h)$.

Meanwhile, the entropy can be exactly evaluated with these analytical PDFs, denoted as $S_\rmth$. These exact results can be used to calibrate our empirical results of $S(\Delta v)$ or $S(h)$, and to help select the optimal bin width or bandwidth.

For this purpose, we choose three analytical PDFs for the experiments. The first is the one-dimensional standard normal distribution:
\begin{equation}
\label{eq:1dgauss}
F(v) = \frac{1} {\sqrt{2\pi}} e^{-\frac{v^2}{2}}, \hskip 5pt -\infty < v < \infty,
\end{equation}
whose $S_{\rmth}=1.419$.

The second experimental PDF is the one-dimensional power-law function:
\begin{equation}
\label{eq:1dpower}
F(v) = 1-\frac{16}{9}v^2,  \hskip 10pt -\frac{3}{4} < v < \frac{3}{4},
\end{equation}
whose $S_{\rmth}=0.2804$. Unlike the first one, this power-law distribution is non-extended for its random variable $v$.

The third one is the three-dimensional isotropic standard normal distribution PDF:
\begin{equation}
\label{eq:3dgauss}
F(\bv) = \frac{1} {(2\pi)^{3/2}} e^{-\frac{\bv^2}{2}}, \hskip 5pt -\infty < \bv < \infty,
\end{equation}
in which $\bv \equiv (v_1,v_2,v_3)$ and $S_{\rmth}=4.257$. This PDF can be further reduced to a one-dimensional distribution, whose PDF is:
\begin{equation}
\label{eq:3dgauss2}
F(v) = \sqrt{\frac{2}{\pi}} v^2 e^{-\frac{v^2}{2}}, \hskip 15pt 0 < v < \infty.
\end{equation}
This reduced one-dimensional distribution, unlike the previous two distributions, is asymmetric in the random variable $v$.

These $S_\rmth$ of the three cases are explicitly indicated in the figures and tables below.

\section{Results}
\label{sec:result}
\subsection{Histogram estimation}
\label{sec:hist}

As described above, PDF of statistical samples can be estimated mainly by histogram and kernel methods. We first present the results by histogram estimation. Figure~\ref{fig:1dgaussDv} shows the experiments with the one-dimensional standard normal distribution of equation~(\ref{eq:1dgauss}).
We do the experiments of the sample size $N$ ranging from $10^3$ to $10^8$, and in every case, we compute $S(\Delta v)$ with different sets of random numbers to give, say $50$, instances. With these $50$ instances, we can derive both the mean $\bar{S}$ and the variance $\sigma$ of the entropy as a function of $\Delta v$. These results are shown in panels from (a) to (f), respectively, and we can see that the variances of all cases are very small. The first derivative of $\bar{S}(\Delta v)$ with respect to $\Delta v$, $\dd \bar{S}(\Delta v)/\dd \ln(\Delta v)$, is also numerically evaluated and shown in the corresponding panels. From this figure, we find that: (1) in all cases, the derivative has a minimum around the cross point of the curve $\bar{S}(\Delta v)$ and the straight line $S_\rmth$, (2) the minimum of the derivative goes to zero with increasing sample size $N$, and (3) $\Delta v_{\dm}$, at which the derivative takes its minimum, approaches the abscissa of the cross point of $\bar{S}(\Delta v)$ and $S_\rmth$, such that $\bar{S}(\Delta v_\dm)$ asymptotically approaches $S_\rmth$.

With exactly the same procedures, we do experiments with another two analytical PDFs, i.e. the one-dimensional power-law and three-dimensional isotropic standard normal distribution, of equations~(\ref{eq:1dpower}) and (\ref{eq:3dgauss}), respectively. The above findings are also applied to these two cases. In Table~\ref{tab1}, we give the relevant means and variances at $\Delta v_{\dm}$ for all the cases above mentioned.

Note that in Figure~\ref{fig:1dgaussDv}, we show the first derivative in the form of $\dd \bar{S}/\dd\ln(\Delta v)$, rather than $\dd \bar{S}/\dd\Delta v$. We can see from the figure, that at $\Delta v_\dm$, $\bar{S}(\Delta v)$ are nearly flat, that is, $\dd \bar{S}/\dd\Delta v \sim 0$. Hence,
\begin{eqnarray}
\frac{\dd^2 \bar{S}}{\dd(\ln\Delta v)^2} {\Big|}_{\Delta v_\dm} & = & \Delta v\frac{\dd \bar{S}}{\dd(\Delta v)}{\Big|}_{\Delta v_\dm} + \Delta v^2\frac{\dd^2 \bar{S}}{\dd(\Delta v)^2} {\Big|}_{\Delta v_\dm}  \nonumber \\
& \approx & \Delta v^2\frac{\dd^2 \bar{S}}{\dd(\Delta v)^2} {\Big|}_{\Delta v_\dm} = 0.  \nonumber
\end{eqnarray}
So the minima of $\dd \bar{S}/\dd\ln(\Delta v)$ and $\dd \bar{S}/\dd\Delta v$ are nearly at the same places.

\begin{figure*}
\centerline{\includegraphics[width=2.05\columnwidth]{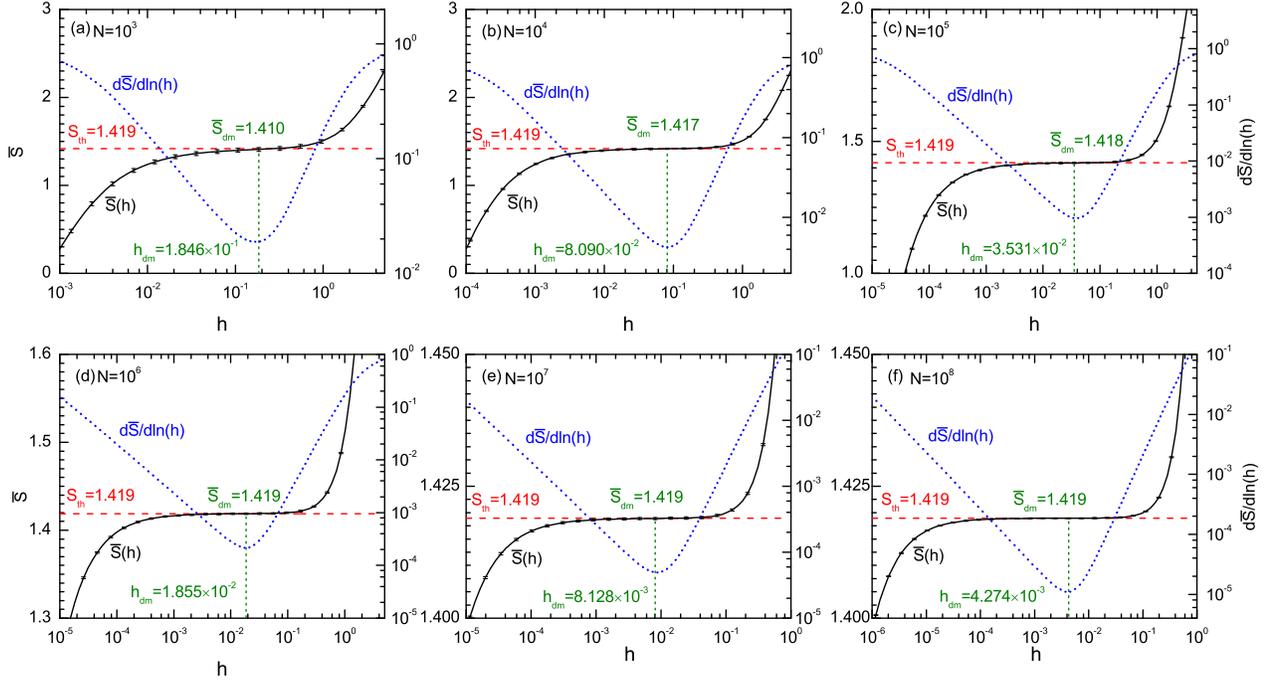}}
\caption{$\bar{S}(h)$, the averaged entropy evaluated with kernel estimation and its first derivative w.r.t to $h$, $\dd \bar{S}/\dd\ln (h)$, for the one-dimensional standard normal distribution of equation~(\ref{eq:1dgauss}). The kernel function is the \citet{epan69} form, and $h$ is the bandwidth of the kernel function. Error bars of $\bar{S}(h)$ curves indicate the standard deviation. The entropies and their derivatives as functions of $h$ are indicated as solid and dotted lines, respectively. Similar to the cases of histogram estimation, in every case, the vertical axis on the left represents $\bar{S}(h)$, and the right represents the derivative. The mean and the standard deviation are evaluated with $50$ instances. Again, the subscript `dm' in the figure indicates `derivative's minimum'.}
\label{fig:1dgaussEpan}
\end{figure*}

\subsection{Kernel estimation}
\label{sec:kernel}

A non-negative real-valued function $K(u)$ is called a kernel function, if satisfying the following two conditions:
\begin{displaymath}
\label{eq:kernel}
\int_{-\infty}^{+\infty}K(u)\dd u =1; {\hskip 20pt}  K(u)=K(-u).
\end{displaymath}
The PDF of a statistical sample, $x_i$, with $i$ running from 1 to the sample size $N$, can be estimated by using the kernel function:
\begin{equation}
\label{eq:funkh}
\hat{F}_h(x) = \frac{1}{h N}\sum\limits_{i=1}^{N}K(\frac{x-x_i}{h}),
\end{equation}
in which $h$ is the bandwidth of the kernel. If the kernel is a derivable function, then $\hat{F}_h(x)$ is also derivable. For this reason, kernel estimation is believed to be superior to histogram method.

The kernel function we used here is \citet{epan69} function $K_e(u)$:
\begin{equation}
\label{eq:epan}
K_e(u) = \frac{3}{4}(1 - u^2),  {\hskip 10pt} -1 < u < 1,
\end{equation}
who has the highest efficiency, in contrast to other kernel functions, to yield the optimal rate of convergence of the mean integrated squared error \citep{silverman86,gentle09}.

We give the results from kernel estimation for one-dimensional normal distribution in Figure~\ref{fig:1dgaussEpan} and the relevant means and variances for all cases at $\Delta v_{\dm}$ in Table~\ref{tab2}. These results are all parallel to the previous results from histograms. Again, we can see that the minimum of the first derivative of $S(h)$ can be regarded as a selector for the optimal bandwidth $h_\dm$. Compared with the histogram estimation, the advantages of kernel estimation are obvious, in that all the curves, $S(h)$ and $\dd S/\dd h$ are smooth and should be derivable.

We also considered other kernel functions, such as uniform and Gaussian \citep{silverman86}. The results are the same as with Epanechnikov (not shown).

\begin{table*}
\caption{Results derived from kernel estimation. All are parallel to those of histogram estimation. For details see Table~\ref{tab1}.}
\label{tab2}
\begin{tabular}{cccccccccc}
\hline
  & \multicolumn{3}{c}{1D normal} & \multicolumn{3}{c}{1D power-law} & \multicolumn{3}{c}{3D normal} \\
N &  $h_{\rm dm}$ & $\bar{S}_{\rm dm}$  & $\sigma(\bar{S}_{\rm dm})$ & $h_{\rm dm}$  & $\bar{S}_{\rm dm}$ & $\sigma(\bar{S}_{\rm dm})$ & $h_{\rm dm}$ & $\bar{S}_{\rm dm}$ & $\sigma(\bar{S}_{\rm dm})$ \\
\hline
$10^3$ &  1.846$\times10^{-1}$ & 1.410  & 2.128$\times10^{-2}$ & 4.213$\times10^{-2}$  & 0.2722 & 1.040$\times10^{-2}$ & 1.539$\times10^{-1}$ &  4.259  & 3.562$\times10^{-2}$ \\
$10^4$ &  8.090$\times10^{-2}$ & 1.417  & 6.634$\times10^{-3}$ & 1.693$\times10^{-2}$  & 0.2781 & 3.273$\times10^{-3}$ & 6.741$\times10^{-2}$ &  4.255  & 1.122$\times10^{-2}$ \\
$10^5$ &  3.531$\times10^{-2}$ & 1.418  & 2.131$\times10^{-3}$ & 6.804$\times10^{-3}$  & 0.2796 & 4.976$\times10^{-4}$ & 3.531$\times10^{-2}$ &  4.257  & 3.990$\times10^{-3}$ \\
$10^6$ &  1.855$\times10^{-2}$ & 1.419  & 8.126$\times10^{-4}$ & 3.281$\times10^{-3}$  & 0.2812 & 2.454$\times10^{-5}$ & 1.546$\times10^{-2}$ &  4.256  & 1.216$\times10^{-3}$ \\
$10^7$ &  8.128$\times10^{-3}$ & 1.419  & 2.040$\times10^{-4}$ & 1.319$\times10^{-3}$  & 0.2805 & 1.327$\times10^{-5}$ & 5.644$\times10^{-3}$ &  4.257  & 3.844$\times10^{-4}$ \\
$10^8$ &  4.274$\times10^{-3}$ & 1.419  & 8.082$\times10^{-5}$ & 7.631$\times10^{-4}$  & 0.2804 & 5.985$\times10^{-7}$ & 2.473$\times10^{-3}$ &  4.257  & 1.240$\times10^{-4}$ \\ \hline
theoretic & -- & 1.419 & -- & -- & 0.2804 & -- & -- & 4.257 & -- \\
\hline
\end{tabular}
\end{table*}

\subsection{Comparison with previous estimators}
\label{sec:comparison}

Prior to our results, there are some well-known estimators for the optimal bin width of histogram or bandwidth of kernel estimation. The \citet{scott79} optimal bin width is one of them,
\begin{equation}
\label{eq:scott}
\Delta v^{*} = \big[ \frac{6}{R(f')}\big]^{1/3} N^{-1/3},
\end{equation}
in which $f'$ is the first derivative of $f$, and the functional $R[g]$ is defined for the roughness of $g$ as
\begin{displaymath}
R(g)=\int_{-\infty}^{+\infty}g(x)^2\dd x.
\end{displaymath}

Another one is the optimal bandwidth for kernel estimation \citep{silverman86,jones96}:
\begin{equation}
\label{eq:hopt}
h_{\rm AMISE}=\big[\frac{R(K)}{R(f'')(\int x^2 K(x)\dd x)^2}\big]^{1/5} N^{-1/5}.
\end{equation}
Note that $h_{\rm AMISE}$ scales with the sample size as $N^{-1/5}$, while $\Delta v^{*}$ scales as $N^{-1/3}$.

The kernel estimation is believed to be superior to the histogram method, and indeed, we notice that $\bar{S}(h)$ as well as $\dd \bar{S}/\dd h$ are much smoother than their counterparts of histogram estimation. So we just concentrate on the kernel estimation below.

\begin{figure}
\centerline{\includegraphics[width=1.0\columnwidth]{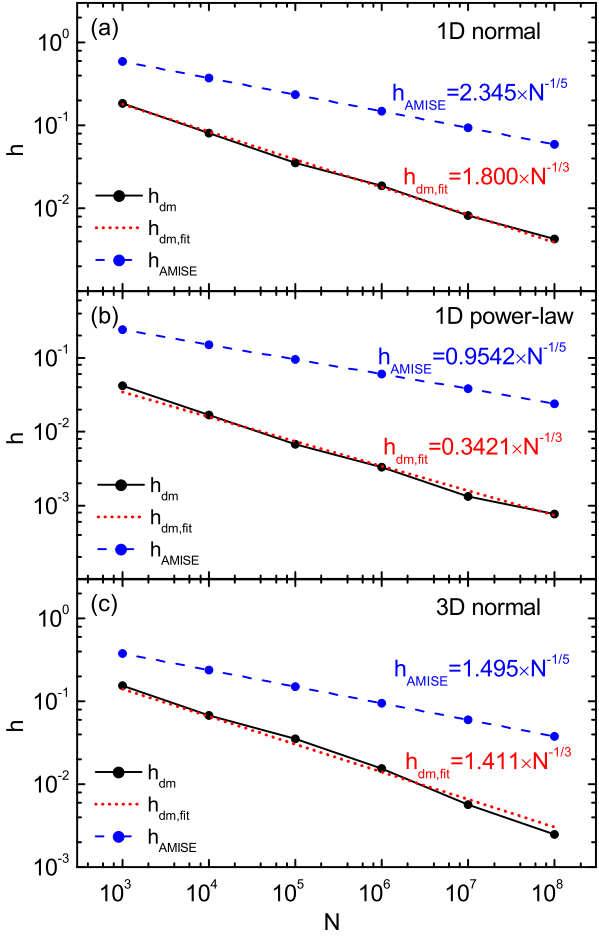}}
\caption{Relation of the bandwidth $h$ and sample size $N$ for the case of kernel estimation. The three panels correspond to the three experimental PDFs of equations~(\ref{eq:1dgauss}), (\ref{eq:1dpower}) and (\ref{eq:3dgauss}). The bandwidth $h_{\rm AMISE}$ that is derived by minimizing AMISE is also shown for comparison.}
\label{fig:hNp}
\end{figure}

Figure~\ref{fig:hNp} shows the relations of $h_{\rm dm}$ with the sample size $N$. We see that in all the three cases, differences between $h_{\rm dm}$ and $h_{\rm AMISE}$ are significant, and $h_{\rm dm}$ well scales as $N^{-1/3}$, in contrast to $h_{\rm AMISE} \propto N^{-1/5}$. It is interesting to note that this $N^{-1/3}$-scaling is similar to that of Scott's formula of equation~(\ref{eq:scott}) for optimal bin width of histogram estimation.

\begin{figure}
\centerline{\includegraphics[width=1.0\columnwidth]{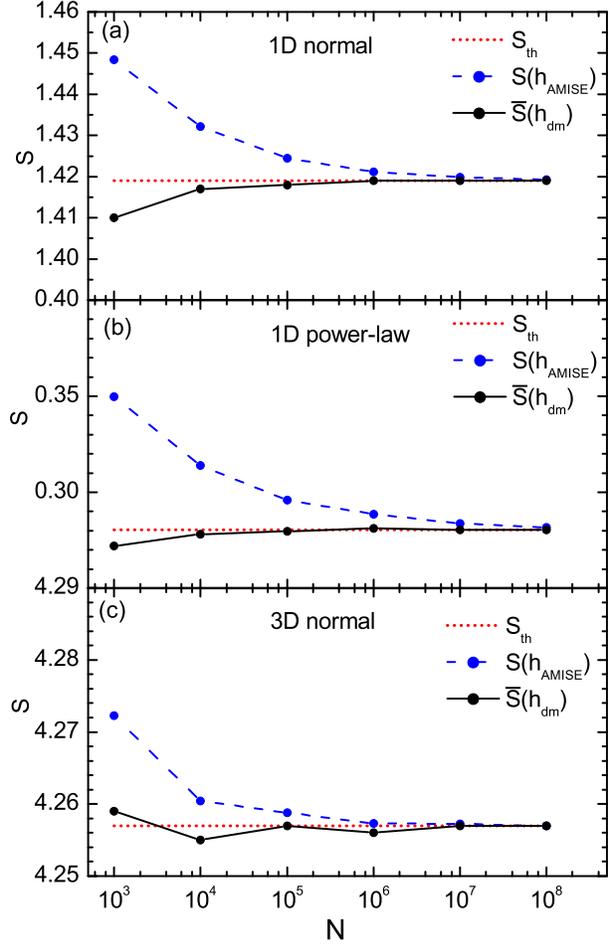}}
\caption{Relation of entropy $S$ and sample size $N$ for the case of kernel estimation. The three panels correspond to the three experimental PDFs of equations~(\ref{eq:1dgauss}), (\ref{eq:1dpower}) and (\ref{eq:3dgauss}). Entropies of $h_{\rm AMISE}$ are also shown for comparison.}
\label{fig:SNp}
\end{figure}

Entropies evaluated at $h_{\rm dm}$ and $h_{\rm AMISE}$ of the three cases are shown in Figure~\ref{fig:SNp}. It can be seen that $\bar{S}(h_{\rm dm})$ is much closer to entropy's true value $S_\rmth$ than $S(h_{\rm AMISE})$, but the difference between the two vanishes asymptotically with increasing sample size. Nonetheless, as far as the statistical computation of entropy is concerned, $h_{\rm AMISE}$, selected by minimizing AMISE, is less than the optimal bandwidth $h_{\rm dm}$.

\subsection{The first derivative of entropy as bandwidth selector}
\label{sec:selector}

From previous results, we have seen that the first derivative of $S(\Delta v)$ or $S(h)$ can be used as a selector for the optimal bin width or bandwidth. We just concentrate on the kernel method here, and present some analytical results. With the estimated PDF $\hat{F}_h$, the first derivative of the entropy is
\begin{eqnarray}
\label{eq:1dmos}
\frac{\dd S(h)}{\dd h}  & = & -\frac{\dd}{\dd h}\int\hat{F}_h\ln\hat{F}_h\dd v \nonumber \\
& = & -\int\big(\frac{\partial \hat{F}_h}{\partial h} \ln \hat{F}_h + \frac{\partial \hat{F}_h} {\partial h} \big) \dd v \nonumber \\
& = & -\int\frac{\partial \hat{F}_h}{\partial h}\ln\hat{F}_h \dd v,
\end{eqnarray}
in which we use the normalization $\int\hat{F}_h\dd v = 1$, and the fact that the partial derivative w.r.t. $h$ can be taken out from the integral.

The second derivative of $S(h)$ is
\begin{eqnarray}
\label{eq:2dmos}
\frac{\dd^2 S(h)}{\dd h^2}  & = & -\int \big[\frac{\partial^2 \hat{F}_h}{\partial h^2} \ln\hat{F}_h + \frac{1}{\hat{F}_h}\big(\frac{\partial \hat{F}_h}{\partial h}\big)^2\big] \dd v \nonumber \\
& = &  -\int \big(\frac{\partial^2 \hat{F}_h}{\partial h^2} \ln\hat{F}_h  - \hat{F}_h \frac{\partial^2 \ln\hat{F}_h}{\partial h^2} + \frac{\partial^2\hat{F}_h}{\partial h^2}\big)\dd v  \nonumber \\
& = & -\int \frac{\partial^2 \hat{F}_h}{\partial h^2} \ln\hat{F}_h \dd v  + \int \hat{F}_h \frac{\partial^2 \ln\hat{F}_h}{\partial h^2} \dd v.
\end{eqnarray}
In deriving the last line of the above relation, we again use the normalization $\int\hat{F}_h\dd v = 1$. Hence, the minimum of the first derivative of $S(h)$ should be picked out by $\dd^2 S/\dd h^2 =0$, and from the above equation, we have
\begin{equation}
\label{eq:sel2dmos}
\int \frac{\partial^2 \hat{F}_h}{\partial h^2} \ln\hat{F}_h \dd v  =  \int \hat{F}_h \frac{\partial^2 \ln\hat{F}_h}{\partial h^2} \dd v.
\end{equation}
We can make use of this relation to pick out the optimal bandwidth $h_{\rm dm}$.

Note that both $\Delta v^{*}$ of equation~(\ref{eq:scott}) and $h_{\rm AMISE}$ of equation~(\ref{eq:hopt}) depend on the unknown true PDF $f$, which is usually difficult or even impossible to acquire. On the contrary, the selector of optimal bandwidth based on the first derivative of entropy is purely data-based, and hence, our approach is much superior to the existing methods.

However, we do not provide a proof on the existence of the minimum of the first derivative of entropy, and do not explain why such a minimum can help pick out the optimal bandwidth, either. These issues are surely of great importance and deserve further investigations.

\section{Summary and discussions}
\label{sec:summy}

Entropy is a very important concept, and in statistical mechanics, it is computed with the probability distribution function of the system under consideration. In our statistical-mechanical investigations of self-gravitating systems, however, we have to do statistical computation of entropy directly from statistical samples, without knowing the underlying analytic PDF. Thus, the evaluation of entropy is actually related to the estimation of the PDF from statistical samples in data-form.

Usually, there are two approaches to estimate a PDF from a statistical sample. The first density estimation method is histogram. Another way is kernel estimation, which is considered to be superior to the histogram. We use both the methods to evaluate the entropy, and we find that the entropy thus computed depends on the bin width of the histogram, or bandwidth of kernel method. Concretely, the entropy is a monotonic increasing function of the bin width or bandwidth. We attribute this monotonicity to the PDF coarse-graining.

Thus, it is a difficulty on how to select an optimal bin width/bandwidth for computing the entropy. Fortunately, we notice that there may exist a minimum of the first derivatives of entropy for both histogram and kernel estimation, and this minimum may correspond to the optimal bin width and bandwidth.

We perform a large amount of numerical experiments to verify this finding. First, we select three analytical PDFs, one-dimensional standard normal, one-dimensional power-law, and three-dimensional standard normal distribution, and with Monte Carlo technique, we draw $N$ random data from these analytical PDFs, respectively. With these statistical samples, we construct the estimator of the true PDFs of both histogram and kernel estimation. Secondly, we compute the entropy with the estimated PDFs, so in this way, we derive the curves of $S(\Delta v)$ or $S(h)$, in which $\Delta v$ or $h$ are the bin width of histogram or bandwidth of kernel, respectively. Meanwhile, the entropy can be exactly evaluated with these analytical PDFs. These exact results can be used to calibrate our empirical results, and to help select the optimal bin width or bandwidth.

We do the same experiments for the three PDFs with the sample size ranging from $10^3$ to $10^8$. In all cases, whatever using histogram or kernel estimation, we find that:
\begin{itemize}
\item The first derivative of entropy indeed has a minimum around the cross-point of the entropy curve $S(\Delta v)$ or $S(h)$ and the theoretical straight line $S_\rmth$.

\item The minimum of the derivative goes to zero with increasing sample size.

\item $\Delta v_{\dm}$ or $h_{\dm}$, at which the derivative takes its minimum, asymptotically approaches the abscissa of the above-mentioned cross-point with increasing sample size, such that the entropy at $\Delta v_\dm$ or $h_{\dm}$ asymptotically approaches the theoretical value $S_\rmth$.

\item $h_{\rm dm}$ scales with the sample size as $N^{-1/3}$, in contrast to $h_{\rm AMISE}$, the bandwidth selected by minimizing AMISE, whose scaling is $\propto N^{-1/5}$.

\item The entropy evaluated at $h_{\rm dm}$ is much closer to the true value $S_\rmth$ than at $h_{\rm AMISE}$, but the difference between the two vanishes asymptotically with increasing sample size.
\end{itemize}

Hence, we see that the minimum of the first derivative of $S(\Delta v)$ or $S(h)$ can be used as a selector for the optimal bin-width or bandwidth of density estimation, and $h_{\rm AMISE}$ is less the optimal bandwidth than $h_{\rm dm}$ for computing entropy.

Note that both Scott's optimal bin width $\Delta v^{*}$ and $h_{\rm AMISE}$ depend on the unknown underlying PDF of the system, which is usually difficult or even impossible to acquire. On the contrary, the estimator of optimal bandwidth selected from the minimum of the first derivative of entropy is purely data-based, and hence, our method is clearly superior to the existing methods.

We emphasize that our results are by no means restricted to one-dimensional, but can also be extended to multivariate cases. Finally, we acknowledge that we do not provide a robust mathematical proof of the existence of the minimum of the first derivative of entropy, nor theoretically explain why such a minimum can help pick out the optimal bandwidth. These issues are surely of great importance and deserve further investigations. We leave these issues with those specialists who are interested in them.

\section*{Acknowledgements}

We thank the referee very much for many constructive suggestions and comments, especially for the reminding of the techniques for estimating information-theoretic quantities developed by \citet{wolpert95} and \citet{wolpert13}. This work is supported by the National Basic Research Program of China (no: 2010CB832805) and by the National Science Foundation of China (no. 11273013), and also supported by the Open Project Program of State Key Laboratory of Theoretical Physics, Institute of Theoretical Physics, Chinese Academy of Sciences, China (no. Y4KF121CJ1).


\label{lastpage}
\end{document}